\documentclass[conference]{IEEEtran}
\usepackage[cmex10]{amsmath}
\usepackage{hyperref}
\usepackage{graphicx}
\usepackage{epstopdf}
\DeclareGraphicsExtensions{.eps}
\usepackage{url}
\begin{document}
\newcommand{\beq}{\begin{equation}}
\newcommand{\eeq}{\end{equation}}
%
%
\title{Minimum Noise Optical Coatings\\ for Interferometric Detectors\\
of Gravitational Waves}
\author{
\IEEEauthorblockN{Maria Principe}
\IEEEauthorblockA{University of Sannio at Benevento, Italy, and INFN\\
Email: principe@unisannio.it}
}
\maketitle
%
%
\begin{abstract}
\boldmath
Coating Brownian noise is the dominant noise term, in a frequency band from a
few tens to a few hundreds Hz, for all Earth-bound detectors of gravitational waves. Minimizing such noise is mandatory to increase the visibility distance
of these instruments, and eventually reach their quantum limit. Several strategies
are possible to achieve this goal. Layer thickness optimization is  the simplest
option, yielding a sensible noise reduction with limited technological challenges.
Experimental results confirm the accuracy of the underlying theory, 
and the robustness of the design.
\end{abstract}
%
%
\section{Introduction}
The birth of  Gravitational Wave Astronomy will open a new and unique window 
on the Universe \cite{GWA}.
Several gravitational wave (henceforth GW) detectors are being constructed, upgraded or planned, including LIGO \cite{1}, GEO \cite{2}, VIRGO \cite{3}, TAMA \cite{4}, ACIGA \cite{5}, INDIGO \cite{indigo}, KAGRA (formerly LCGT) \cite{kagra}, and ET \cite{ET}, in an unprecedented multi-national effort. 
Gravitational waves, predicted by Einstein relativistic theory of gravitation \cite{GR}, are ripples in the spacetime fabric produced by  massive cosmic objects in accelerated motion, which can be detected using very long baseline optical interferometers \cite{Saulson}. 
The sensitivity of these instruments is limited by noises of different origin (e.g., seismic, thermal and quantum, see Figure 1). 
A reduction of the noise floor level by a factor $p$ entails a  $p^{-3/2}$ boost of the 
instrument visibility volume \cite{Saulson}. 
For Earth bound detectors, the power spectral density of the noise floor is minimum in a frequency band between a few tens and a few hundreds Hz,  where several cosmic sources of gravitational waves (henceforth GW) are deemed to exist \cite{GWA},  and  Brownian noise in the highly reflective coatings of the test masses making up the end-mirrors of the interferometer arms is the dominant noise source \cite{Harry2006}.
Minimizing coating Brownian noise is a must to reach (and eventually beat \cite{SSI}) the sensitivity quantum limit.\\

In this paper I review  possible approaches to coating noise minimization,
with special emphasis on coating design optimization, which was invented  at the University of Sannio \cite{TheBook} and successfully implemented in collaboration with the Laboratorie des Mat\'{e}riaux Avanc\'{e}s, Lyon (FR), and the LIGO Lab of the California Institute of Technology,  Pasadena CA (USA) \cite{PRD}.
%
\begin{figure} [!h]
\centerline{\includegraphics[width=10.5cm]{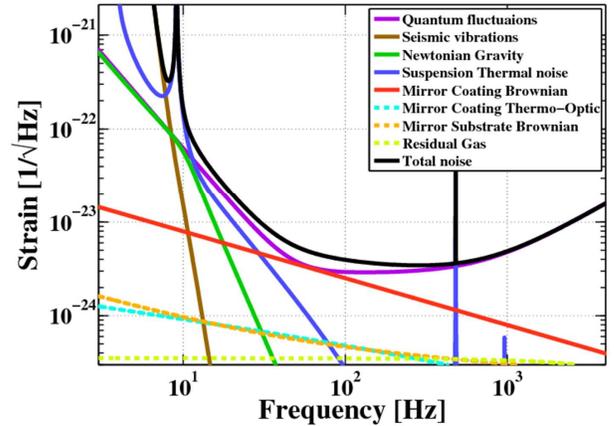}}
\vspace*{-6.5cm}
\caption{Noise power spectral density budget of the advanced LIGO detector in strain (gravitational wave amplitude) units. } 
\label{fig_noises}
\end{figure}
%
%
\section{Coating Thermal Noise}
\label{sec:TN}
%
Using the fluctuation-dissipation theorem,  
the Brownian noise power spectral density 
in the interferometer test-mass mirror coatings 
can be cast in the form \cite{Harry2006}
\beq
S_B(f) = \frac{2 k_B T}{\pi^{3/2} f} \frac{(1-\sigma_s^2)}{w Y_s } 
\phi_{c},
\label{eq:SBf}
\eeq
where $k_B$ is Boltzmann's constant, 
$T$  the absolute temperature, 
$w$ the half-width of the (Gaussian) laser beam, 
$\sigma_s$  and $Y_s$ are 
the Poisson's ratio  and Young's  elastic modulus of the substrate, 
and  $\phi_{c}$ is the effective mechanical loss angle of  coating .
To reduce $S_B$ one thus could: 
i) cool the mirrors (i.e., decrease $T$); 
ii) increase the illuminated area (represented in \ref{eq:SBf} by $w$);
iii) reduce the coating loss angle $\phi_{c}$.
I shall focus here on the third option (the other two will be 
shortly discussed in Section \ref{sec:more}).\\
Coatings are presently made of alternating layers 
of two dielectric materials (amorphous glassy oxides)
with different refractive indexes.
In the limit where the materials' Poisson's ratios are vanishingly small,
we have a simple formula for the coating loss angle \cite{Harry2006}:
\beq
\phi_{c} = b_L d_L \phi_L + b_H d_H \phi_H,
\label{eq:PhiEff}
\eeq
where $d_L$ and $d_H$ are the total thicknesses 
of the low(er) and high(er) index materials, respectively,
$\phi_{L,H}$ their mechanical loss angles, and
\beq
b_{L,H} = \frac{1}{\sqrt{\pi}w} \left(  \frac{Y_{L,H}}{Y_s}+\frac{Y_s}{Y_{L,H}} \right) ,
\label{eq:bLH}
\eeq
$Y_{L,H}$ denoting the materials' Young's moduli. 
The quantities 
\beq
\gamma_{L,H} \!=\! b_{L,H}\phi_{L,H}
\label{eq:specphi}
\eeq
will be henceforth referred to 
as the {\it specific} loss angles (loss angles per unit thickness).\\
The quarter-wavelength (or Bragg) coating design, where
the thickness of the individual layers is
$\delta_{L,H} \!=\! \lambda_0/4n_{L,H}$,
$\lambda_0$ being the operating wavelength, 
yields the minimum number of layers 
to achieve a prescribed transmittance \cite{TheBook},
and is the usual choice for all applications where noise is not an issue.\\
Careful material downselection led to the choice of  $SiO_2$ (Silica)
and $Ta_2O_5$ (Tantala) as the best available materials 
for the highly reflective (henceforth HR) 
coatings of GW detectors \cite{downselect}, yielding the best tradeoff among  
high dielectric contrast (large ratio $n_h/n_L$), 
low optical absorption (small extinction coefficients, $\kappa_{L,H}$),
and low thermal noise (small specific loss angles $\gamma_{L,H}$).\\
The specific loss angles of Silica and Tantala, however, are quite
different, being $\gamma_H \! \approx \! 10 \gamma_L$. 
This suggests trying {\it different} (non-Bragg) designs, where
{\it the coating loss angle} (hence the coating noise) {\it is minimized 
subject to a coating transmittance constraint}.\\ 
%
\section{Coating Thickness Optimization}
%
Genetic optimization, where {\it no} a-priori assumption is made about
the geometric structure of the sought optimal coatings
(total number of layers and thicknesses of the individual layers) 
showed that  (except for the coating top and bottom layers),
the optimal coating  consists of a stack of {\it equal} doublets 
with optical thicknesses  $z_{L,H}$  such that  $z_L\!+\!z_H \! \approx \!1/2$,
and $z_H \! <\! 1/4 \! <\! z_L$  \cite{genetic}.\\ 
The number of free design parameters is accordingly reduced
to just four: the total number  $N_d$ of doublets,  a quantity $\xi \! \in \! (0,1/4)$,
such that  $z_{L,H}\!=\!1/4 \! \pm \! \xi$, and the optical thicknesses of the
top and bottom layers, $z_T$ and $z_B$.\\
Coating optimization is accordingly most easily implemented
{\it sequentally} through the following steps \cite{TheBook}:
i) start from the quarter wavelength design with (power) transmittance $\tau_0$ 
closest to the design value, and consisting of  $N_d=N_0$ doublets;
ii) add one doublet, and adjust the layers’  thicknesses (by varying the single parameter $\xi$) to make the coating transmittance equal to $\tau_0$; 
iii) calculate the loss angle $\phi_c$, and repeat step ii) until $\phi_c$ reaches a minimum.
This procedure results into a {\it shallow} minimum, as shown in Figure 2,
suggesting that the design will be robust against possible  inaccuraccies
in the assumed values of $\gamma_{L,H}$, and unavoidable coating deposition tolerances \cite{TheBook}. 
%
\begin{figure} [!h]
\centerline{\includegraphics[width=9.5cm]{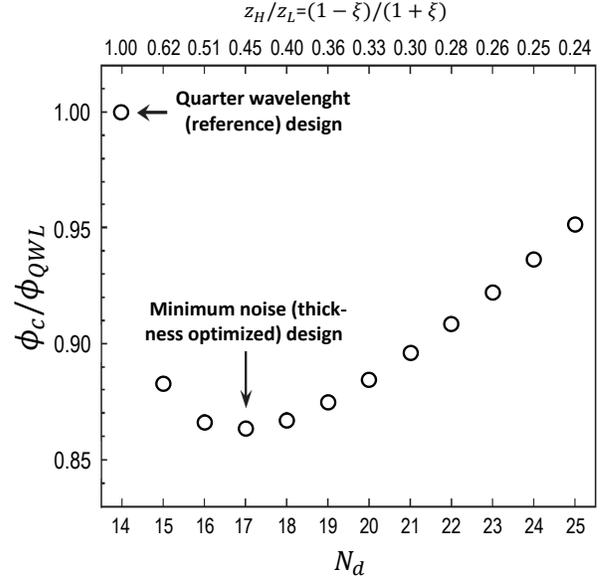}}
\vspace*{-4.0cm}
\caption{Loss angle (normalized to that of the reference quarter-wavelengh
design) of Silica/Tantala coatings with identical transmittance (here $287ppm$) but different number $N_d$ of doublets, and different layer thicknesses $z_{L,H}$. The quarter wavelength and minimum noise (optimized thickness) designs are indicated.} 
\label{fig_OPT}
\end{figure}
%
The final steps consists in:
iv) adjusting the thickness $z_B$ of the bottom (H)-layer to minimize noise, and  
v) adjusting the thickness $z_T$ of the top (L)-layer to bring back the transmittance to $\tau_0$.
%
\section{Optimized Coating Prototypes}
%
The above coating optimization procedure was used to produce 
a batch of mirrors suited for the  Caltech {\it Thermal Noise Interferometer} (TNI), 
an instrument designed for the {\it direct}  measurement of  coating thermal noise \cite{Black}, shown in Figure 3.\\
%
\begin{figure} [!h]
\centerline{\includegraphics[width=9.5cm]{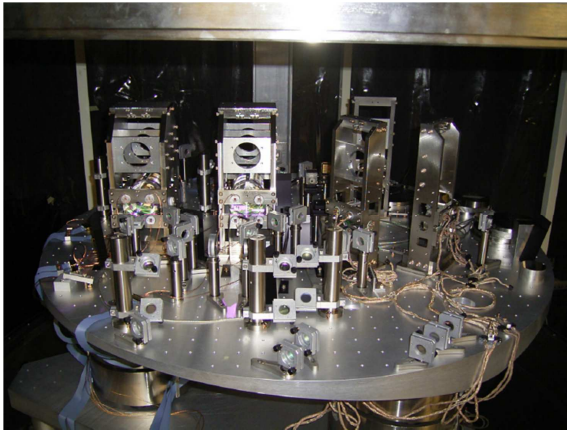}}
\vspace*{-5.5cm}
\caption{The Caltech Thermal Noise Interferometer with its vacuum dome
lifted (courtesy E. Black). } 
\label{fig_TNI}
\end{figure}
%
The optimized prototypes were designed at the University of Sannio, and
manufactured by LMA (Laboratoire Materiaux Avanc\'{es} of CNRS-In2P3,
Lyon, FR), using their large ion-beam sputtering (IBS) chamber, shown in Figure 4.\\
%
\begin{figure} [!h]
\centerline{\includegraphics[width=9.5cm]{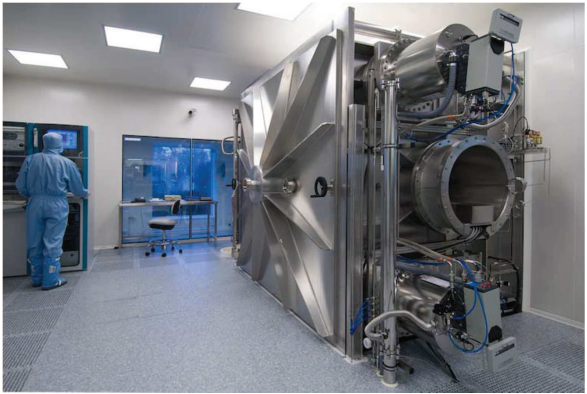}}
\vspace*{-5.5cm}
\caption{The large IBS coater at the Laboratoire des Materiaux
Avanc\'{e}s  (LMA) of the CNRS-IN2P3, Lyon FR, were the optimized 
coating prototypes designed for the TNI were manufactured (courtesy
LMA). } 
\label{fig_LMA}
\end{figure}
%
The optimized coating thermal noise was measured with high accuracy,
and compared to that of standard quarter-wavelength coatings having 
the {\it same} transmittance ($\tau\!=\!287\mbox{ ppm } @ 1064 \mbox{ nm}$). \\
The optimized and reference (quarter-wavelength) coatings are sketched
in Figure 5. The measurement setup, and the data analysis procedure are described in
detail in \cite{PRD}.
%
\begin{figure} [!h]
\centerline{\includegraphics[width=9.5cm]{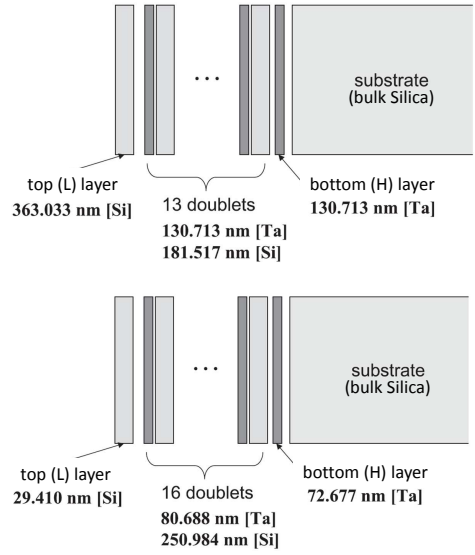}}
\vspace*{-3.5cm}
\caption{Structure of the reference (quarter-wavelength, top) 
and thickness optimized (bottom) TNI coating prototypes.} 
\label{fig_proto}
\end{figure}
%
The measured  loss angle of the optimized coatings was lower by a factor
$p=0.82\pm0.04$ compared to that of the quarter wavelength coatings.
This value reproduced, within the estimated uncertainty range of the measurements
and the nominal accuracy of the material parameters, our modeling predictions, 
confirming the validity and effectiveness of our optimization strategy.
%
\subsection{Optimized Dichroic Coatings}  
%
Advanced (2nd generation) interferometers
will use the $2$nd harmonic of the laser beam 
for alignment purposes. 
The test mass coatings must be accordingly {\it dichroic},
and besides being highly reflective 
at the fundamental wavelength $\lambda_0$ 
(with typical transmittances of a few ppm)
should provide some reflectance 
also at $\lambda_1 \!=\! \lambda_0/2$
(with typical values around $0.9$).\\
The originally proposed (reference) dichroic design
for the AdLIGO coatings consists  
of a  stack of $N_1$ doublets grown on top of the mirror substrate, 
with geometrical  thicknesses $\delta_{L,H}$ such that 
\beq
n_H(\lambda_1)\delta_H \!=\! \frac{\lambda_1}{4}, 
\mbox{  } 
n_L(\lambda_1)\delta_L \!=\! \frac{3\lambda_1}{4},
\label{eq:botstack}
\eeq
topped by a second stack of $N_0$ doublets 
with geometrical thicknesses $\Delta_{L,H}$ such that
\beq
n_H(\lambda_0)\Delta_H \!=\! n_L(\lambda_0)\Delta_L 
\!=\! \frac{\lambda_0}{4}. 
\label{eq:topstack}
\eeq
Neglecting chromatic dispersion in the materials, i.e., assuming 
$n_{L,H}(\lambda_0) \!=\! n_{L,H}(\lambda_1)\!=\!n_{L,H}$, 
eq. (\ref{eq:topstack}) entails 
\beq
n_H\Delta_H \!=\! n_L\Delta_L \!=\! \frac{\lambda_1}{2}.
\eeq
Hence, at $\lambda \!=\! \lambda_1$ the top stack is transparent, 
and the bottom stack, which is effectively quarter-wavelength,
is designed to provide the prescribed reflectance.
Similarly, eq. (\ref{eq:botstack}) entails
\beq
n_H\delta_H \!=\! \frac{\lambda_0}{8}, 
\mbox{    } 
n_L\delta_L \!=\! \frac{3\lambda_0}{8}.
\eeq
Hence, at $\lambda\!=\!\lambda_0$,
the bottom stack contributes part of the required reflectance,
and the top stack, which is quarter-wavelength, 
is designed to bring the reflectance to the prescribed level \cite{dichroic1}.\\
In order to shed light on the structure of minimal noise dichroic coatings,
without making any a-priori assumption about the number and thickness 
of the individual layers, we resorted again to genetic optimization  
to seek  coating configurations which minimize the coating Brownian noise 
under a {\it dichroic} transmittance constraint \cite{dichroic1}.\\
Genetically optimized coatings were found to consist of a stack of equal doublets
(except for the  coating top and bottom layers) with thicknesses $z_{L,H}$
such that $z_H \! <\! 1/4 \! <\! z_L$ at $\lambda \!=\! \lambda_0$,  like
for single-wavelenght operation.  At variance with this latter, however,
in the dichroic case $z_L\! +\!z_H \neq\! 1/2$ .\\
The number of free design parameters is accordingly reduced to {\it five}:
the total number  $N_d$ of doublets,  the quantities $\xi_{L,H} \! \in \! (0,1/4)$,
such that $z_L \!=\!1/4 \! +\! \xi_L$ and  $z_H \!=\!1/4 \! -\! \xi_H$
at $\lambda \!=\! \lambda_0$,  and the optical thicknesses $z_T$ and $z_B$ 
of the  top and bottom layers.\\
The optimization strategy in the dichroic case  reduces to the following \cite{dichroic2}: 
i) find by trial-and-error the {\it minimum} value of $N_d$ (the number
of doublets) for which the region $\Sigma(N_d)$ in the ($\xi_L$, $\xi_H$) 
plane where  the dichroic transmittance requirements are statisfied is {\it not} empty;
ii) identify  the point  $\{ \xi_L^*, \xi^*_H\} \! \in\! \Sigma(N_d) $  
where the coating loss angle is minimum, let it be  $\phi^*_c(N_d)$;
iii) increase $N_d$ and repeat step ii) 
until  a minimum of $\phi^*_c(N_d)$ is reached.\\
The thickness $z_B$ and $z_T$ of the bottom (H) and top (L) layer
can be adjusted for further noise reduction, similar to the
single wavelength case, or to enforce additional requirements 
(e.g., to minimize the electric field intensity on
the coating face, to prevent dust contamination).
The shape of the  $\Sigma(N_d)$ region in the $(\xi_L,\xi_H)$ plane is
sketched in Figure  6.\\
\vspace*{-1.5cm}
%
\begin{figure} [!h]
\centerline{\includegraphics[width=13.5cm]{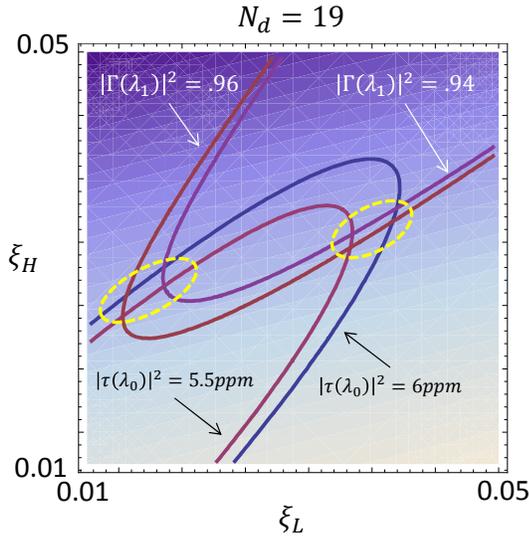}}
\vspace*{-9.cm}
\caption{Constant transmittance/reflectance loci in the $(\xi_L,\xi_H)$ plane for a
$19$-doublets Silica/Tantala coating. The $\Sigma(N_d)$ region for dichroic response constraints of the interval type (AdLIGO) consists of two disjoint subsets (highlighted by the dashed yellow loops), which collapse into two distinct points in the case of equality constraints. Darker/lighter blue shades indicate higher/lower Brownian noise levels.} 
\label{fig_dicro1}
\end{figure}
%
\newline
Coating thickness optimization of the AdLIGO end-test-mass (ETM) dichroic
coatings reduces the  loss angle by a factor $\sim 0.88$ compared to the
reference design. The spectral response is also improved, as shown in 
Figure 7. 
Modeling predictions have been fully confirmed 
by TNI measurements on thickness optimized dichroic prototypes
\cite{dichroic3}.
%
\begin{figure} [!h]
\centerline{\includegraphics[width=9.5cm]{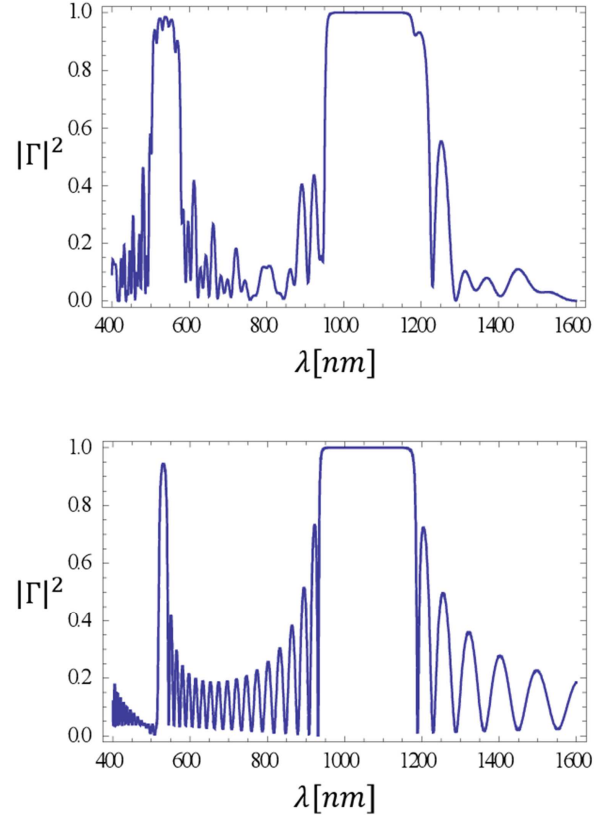}}
\vspace*{-0.35cm}
\caption{Spectral response of reference (top) and thickness optimized (bottom)
dichroic coating for the LIGO end-test-mass  (ETM) mirror. .} 
\label{fig_dicro2}
\end{figure}
%
\section{More Coating Noise Reduction Strategies}
\label{sec:more}
%
In this Section we present a compact overview of  different
test-mass Brownian noise reduction strategies proposed so far, 
including those mentioned at the end of Section \ref{sec:TN} .
%
\subsection{Better Coating Materials}
%
\vspace*{-.35cm}
Coating Brownian noise can be reduced by acting 
on the relevant material properties, 
represented by the specific loss angles $\gamma_{L,H}$ 
in eq. (\ref{eq:specphi}),
and the refraction indexes $n_{L,H}$. 
Smaller $\gamma_{L,H}$ values and larger (contrast) $n_H/n_L$ ratios 
(which help reducing the number of layers needed to achieve a prescribed
reflectance, and hence the coating ticknesses $d_{L,H}$)
yield lower Brownian noise.\\ 
The most successful attempt in this direction is likely represented 
by the development of  $TiO_2::Ta_2O_5$ (Titania-Tantala) 
\cite{dopedTa}  and  $TiO_2::SiO_2$  (Titania-Silica)  \cite{CSIRO}
mixtures. \\
Mechanical losses in amorphous materials are associated with thermally
activated local transitions between the minima of asymmetric bistable potentials,
and can be computed from knowledge of (the distributions of) 
their relevant parameters \cite{Gilroy}.  
Modeling efforts to deduce these latter from first principles are ongoing 
\cite{HaiPing}, \cite{Bassiri}. 
Present knowledge, however is not sufficient for engineering glassy oxide
mixtures with prescribed properties, nor even for improving them,
and the quest for better coating materials still relies
on extensive trial-and-error \cite{Flaminio}.
%
\subsection{Low Temperatures}
%
Lowering the temperature
does not reduce coating Brownian noise 
as much as one would expect from (\ref{eq:SBf}). 
Most coating materials, including Silica and Tantala  
(plain as well as $TiO_2$-doped) 
exhibit a mechanical-loss peak at some low temperature 
(see Figure 8 and \cite{Iain1}),  
whose height and width depend on the material composition, 
and the post-deposition annealing schedule \cite{Iain2}.
Hafnia ($HfO_2$) and Titania ($TiO_2$) are notable
exceptions \cite{cryoHafnia}, \cite{cryoTitania}.
Unfortunately, both are prone to the formation of crystallites 
during the annealing phase, which spoils severely
their optical (scattering) and mechanical (loss angle)
qualities.\\
This difficulty can be circumvented by doping
(co-sputtering) these materials with good glass formers, 
like Silica \cite{Si_buffer}.  This has been
demonstrated to suppress crystallite formation
during high temperature annealing
in $HfO_2$ and $ZrO_2$  \cite{Si_Doped_Hf_Zr}, 
as well as $TiO_2$ \cite{Si_Doped_Ti},
at dopant concentrations $\sim 20\%$.\\
Remarkably, Silica doping does {\em not} affect
the nice low-temperature properties of Hafnia
\cite{Cryo_Si_Hf_OK}. Cryogenic
measurements on Silica doped Titania are
underway \cite{Cryo_Chao}.
\vspace*{-0.5cm}
%
\begin{figure} [!h]
\centerline{\includegraphics[width=8.5cm]{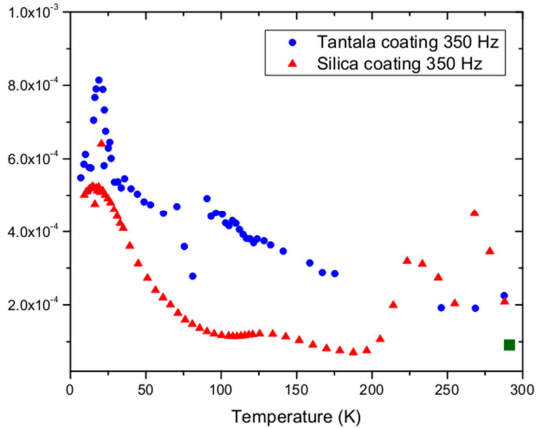}}
\vspace*{-4.cm}
\caption{Mechanical losses vs temperature, from \cite{Iain1}.
Losses increase upon reducing temperature, peaking at a certain
temperature.} 
\label{fig_Cryo}
\end{figure}
%
\subsection{Nanolayered Materials}
%
A possible alternative to co-sputtered glassy mixtures 
is offered by nm-layered materials. 
Their macroscopic properties are amenable to simple 
(effective medium  theory based) modeling \cite{Pinto_EMT},
which makes them easily engineerable.
Nanometer-scale layered films 
of $HfO_2$ and $Al_2O_3$ 
do not crystallize upon annealing
up to very high temperatures \cite{Hf_Al_nanolayered}.
The same applies to nm-layered $TiO_2$ and $SiO_2$ films
\cite{Ti_Si_nanolayered}, \cite{Ti_Si_nanolayered2},
as illustrated in Figure 9.
Cryogenic measurements of the loss angle
of  Hafnia/Silica and  Titania/Silica  nm-layered mixture 
are underway \cite{Cryo_Chao}.
%
\begin{figure} [!h]
\centerline{\includegraphics[width=9.5cm]{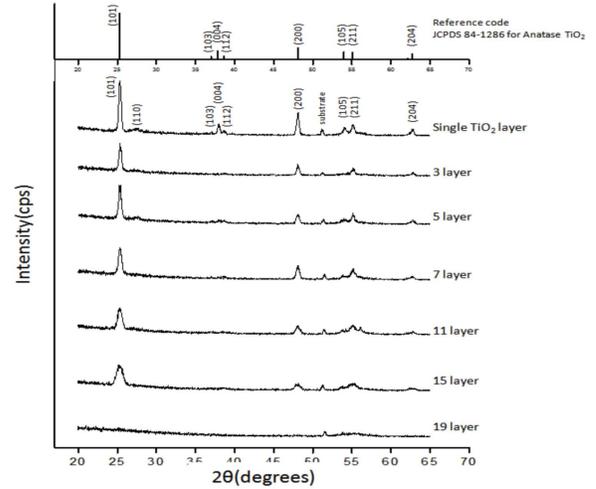}}
\vspace*{-4.5cm}
\caption{X-ray diffraction spectra of different Titania/Silica nm-layered films,
after 24h annealing $@$ 300C. All films are designed to have the same refraction index and optical thickness, but differ in the total number and thickness of the individual layers. As the number of layers increases (and the layer thickness decrases)
the diffraction peak signaling crystallization (Anatase formation) gradually disappears
\cite{Ti_Si_nanolayered2}.} 
\label{fig_XRD}
\end{figure}
%
\subsection{Wide Beams}
%
Wide beams are effective in reducing coating noise 
by averaging out thermal fuctuations
of the mirror surface over a larger illuminated area.
Different families of "wide beams" have been proposed so far,
including "mesa"  \cite{MesaBeam},  hyperbolic \cite{GaldiBeam}  
and Bessel beams \cite{Bondarescu}.
See \cite{PierroBound} for a broad discussion.\\
Gauss-Laguerre modes, in particular, received considerable attention,
since they may fit standard spherical-mirror cavities \cite{HOGL}, 
although imposing much tighter mode-matching and astigmatism  requirements \cite{HOGL2}.
%
\subsection{Radical Alternatives}
%
A number of radical alternatives to present day mirrors, 
based on amorphous glassy oxide dielectric coatings, have also been proposed.
Among these: replacing the mirrors with anti-resonant cavities 
obtained by  leaving only a few coating layers on the front face 
of transparent test masses, and placing the remaining ones on the back face 
(Khalili etalons \cite{Gurko});
adopting non-diffractive, coating-free mirrors, based on total internal reflection 
and Brewster angle coupling \cite{Gossler}; 
using diffractive (grating-based) monolithic (e.g., Silicon or Sapphire) 
mirrors \cite{Bunkoski}; taking advantage of the extreme low
losses of epitaxially grown single-crystalline (e.g., $GaP/AlGaP$) 
coatings \cite{Cole}.
All these concepts hold significant potential and are being actively explored, 
but each of them faces specific  technological and/or conceptual problems which hinder,
at present, their immediate full-scale applicability to  GW detectors.\\
%
\section{Conclusions}
%
Coating design optimization for thermal (Brownian) noise minimization in the test-mass mirrors of interferometric detectors of gravitational waves has been reviewed, with emphasis on geometric (thickness) optimization, which was invented at the University of Sannio,  and developed in Collaboration with LMA and Caltech.\\
Among all test-mass Brownian noise reduction techniques proposed so far, coating thickness optimization is by far the simplest, best understood, technologically less demanding, and cheapest option, capable of reducing the coating noise power spectral density level by a factor $\sim 0.8$, and correspondingly boosting the instrument's visibility distance by a round $\sim 30\%$.
%
\section*{Acknowledgments}
%
The work described in this paper has been done in collaboration with Akira E. Villar, Eric D. Black, Riccardo DeSalvo, Kenneth G. Libbrecht, Christophe Michel, Nazario Morgado, Laurent Pinard, Innocenzo M. Pinto, Vincenzo Pierro, Vincenzo Galdi and Ilaria Taurasi, and has been sponsored in part by the NSF (under the Cooperative Agreement PHY-0757058) and the INFN (under the  "COAT" and "MIDI-BRUT" grants). 
%
%

\end{document}